\documentstyle[12pt]{article}
\begin{document}
\addtolength{\baselineskip}{1mm}
\font\mb=msbm10
\font\helv=cmssbx10

\begin{center}
\begin{flushright}
THU-94/05
\end{flushright}

\vspace{1cm}

{\Large{\bf
Chaotic Scattering Theory of Transport\\[1ex]
and Reaction-Rate Coefficients }}\\
\vspace{1.7cm}
{\large J. R. Dorfman}\\
\vspace{0.3cm}
{\it Institute for Physical Science and Technology and\\
Department of Physics\\
University of Maryland\\
College Park, MD 20742, USA
\\[4ex]
and
\\[4ex]
{\large P. Gaspard}\\
\vspace{0.3cm}
Facult\'e des Sciences and \\
Centre for Nonlinear Phenomena and Complex Systems\\
Universit\'e Libre de Bruxelles, Campus Plaine, Code Postal 231\\
B-1050 Brussels, Belgium}\\
\end{center}

\vspace{4.0cm}

\noindent{PACS numbers:}

\vskip 0.5 cm

\noindent 05.45.+b.   {\it Theory and models of chaotic systems}

\vskip 0.2 cm

\noindent 05.60.+w.   {\it Transport processes: Theory}

\vskip 0.2 cm

\noindent 05.40.+j.   {\it Fluctuation phenomena, random processes, and
Brownian motion}

\vfill\eject
\addtolength{\baselineskip}{10pt}

\centerline{\sc Abstract}
\vskip 0.5 cm
\begin{quote}

The chaotic scattering theory is here extended to obtain escape-rate
expressions for the transport coefficients appropriate for a simple
classical fluid, or for a chemically reacting system. This theory allows
various
transport coefficients such as the coefficients of viscosity, thermal
conductivity, etc., to be expressed in terms of the positive Lyapunov exponents
and Kolmogorov-Sinai entropy of a set of phase space trajectories that take
place on an appropriate fractal repeller. This work generalizes the previous
results of Gaspard and Nicolis for the coefficient of diffusion of a particle
moving in a fixed array of scatterers.
\end{quote}

\vfill\eject

\section{Introduction}

One of the remarkable results of recent studies of non-equilibrium
processes taking place in large systems has been the establishment of new
connections between non-equilibrium statistical mechanics, and dynamical
system theory \cite{EckRue85,Sin91}. Although much more work needs to be done
before these connections are clear and well established, some of the new
results
establish a direct connection between quantities of interest for the
statistical
mechanics of irreversible processes, such as transport coefficients,
and quantities of interest for the dynamical description of the same
system, such as Lyapunov exponents, and Kolmogorov-Sinai entropies. Here we
will restrict our discussion to classical systems for which a reasonable level
of
understanding has been achieved.

One method for developing such connections, which has been developed by Hoover,
Posch and coworkers \cite{HooPos87}, and by Evans, Cohen, Morris and coworkers
\cite{EvaCohMor90}, is based on a study of  systems that are maintained in
non-equilibrium steady states by means of specially constructed external or
internal ``forces'' and a thermostat that removes energy generated in the
system
by the special forces. This method leads to interesting connections between
transport coefficients, such as the coefficients of diffusion or viscosity, and
the Lyapunov exponents of the thermostatted, forced system. For the case of a
thermostatted single particle moving (in two dimensions) in a periodic Lorentz
gas under the action of an electric field, the connection between the diffusion
coefficient, or the electrical conductivity, and the (two) Lyapunov exponents
was
studied by Baranyi, Evans, and Cohen \cite{BarEvaCoh93}, and rigorous results
establishing this connection were obtained by Chernov, Eyink,  Lebowitz and
Sinai \cite{CheEyiLebSin93}. One feature of this approach is that a Hamiltonian
description of the  system is not possible, due to the introduction of the
thermostat and the question naturally arises as to whether a connection of the
sort described is possible for purely Hamiltonian systems.

This question was answered in the affirmative, for the case of a classical
particle moving in an environment of fixed scatterers, by Gaspard, Nicolis,
and coworkers \cite{GasNic90,GasBar91}. They were able to show that the
diffusion
coefficient of the moving particle could be obtained in terms of the escape
rate
for the particle from a bounded region of the scatterers. This escape rate, in
turn, is obtained -- following arguments of Kantz and Grassberger
\cite{KanGra85},
Bohr and Rand \cite{BohRan87}, Tel and coworkers \cite{Tel}, Grebogi, Ott, and
Yorke \cite{GreOttYor88} -- in terms of the positive Lyapunov exponents and
KS-entropy that characterize the set of orbits of the moving particle that are
trapped forever in the bounded region occupied by the scatterers. This set of
``trapped'' orbits form a fractal set of trajectories of the moving particle,
and is referred to as the ``repeller'' \cite{KadTan84}.  The relation between
the
diffusion coefficient, $D$, for the moving particle and the dynamical
quantities
on the repeller is given by \cite{GasNic90}

$$D\ =\ \lim_{L\rightarrow\infty}\  \biggl({L\over{\pi}}\biggr)^2
\Biggl\lbrack \sum_{\lambda_i > 0}\ \lambda_i (L)\ -\ h_{\rm
KS}(L)\Biggr\rbrack
\ . \eqno(1.1)$$

\noindent
Here, we suppose that the scatterers are confined to a  slab of width $L$
in one dimension, and of infinite length in perpendicular spatial directions;
$\sum_{\lambda_i > 0} \lambda_i (L)$ is the sum over all {\em positive}
Lyapunov exponents for trajectories on the repeller, and $h_{\rm KS}(L)$ is
the KS-entropy for trajectories on the repeller. We assume, usually
without proof, that the system has ergodic properties such that the
$\lambda_{i}$ do not depend on the particle trajectory on the repeller.
We also note that for the open systems considered here (i. e. particles not on
the repeller escape from the region containing the scatterers), the KS-entropy
on the repeller does {\em not} equal the sum of the positive Lyapunov
exponents, and the  difference is of order $L^{-2}$ for large $L$, if the
diffusion coefficient exists. The escape-rate expression for the diffusion
coefficient has been evaluated explicitly for a two-dimensional periodic
Lorentz gas \cite{GasBar91}, for a two-dimensional multibaker transformation
\cite{Gas92}, and for a variety of one-dimensional maps \cite{KlaJRD94}.

In the preceding discussion, the diffusion process is actually considered as a
chaotic scattering occuring in a large but finite system.  Several recent works
have been devoted to chaotic scattering and have elucidated the role of the
fractal repeller in this phenomenon \cite{Chaos93}.  Usually, chaotic
scattering
is envisaged on a small scatterer containing a few scattering centers
\cite{Tel,Chaos93,GasRic89}.  However, it was shown that chaotic scattering
becomes controlled by diffusion when the scatterer becomes large enough.
Remarkable relations then exist between the properties of scattering and those
of diffusion, which have recently been described in some detail \cite{Gas94}.

It is the purpose of this paper to extend the chaotic scattering
theory and, in particular, the escape-rate formalism to include other transport
coefficients, and to include a treatment of chemical reaction-rate
coefficients.
This is accomplished by showing that the basic ideas used to derive the
expression for the diffusion coefficient of a moving particle can be easily
extended to apply to other transport and reaction-rate coefficients as well.

In the next Secs. 2 and 3, we present the derivation of the escape-rate
expressions for transport and reaction rate coefficients. This
derivation is based on the fact that all of these coefficients can be related
to
the average mean square displacement of some appropriate dynamical quantity. In
Sec. 4 of the paper, we outline some outstanding and interesting  problems
related to this work. A careful discussion of the theory for chemical
reaction rates is given in Appendix A. Derivations of the relation between the
escape rate of a dynamical quantity from a bounded region and the Lyapunov
exponents and KS-entropy has been given in the literature for systems of a
small
number of dimensions \cite{EckRue85,KanGra85,BohRan87,Tel,GreOttYor88}.  In a
separate paper \cite{GasJRD94}, we will review this derivation, so as to extend
it to systems with many degrees of freedom, to systems that may be of
nonhyperbolic type, and to illustrate the method with applications to lattice
gas automata and to hard sphere gas systems.

\section{Transport Coefficients and their Helfand Moments}

We begin by constructing a large system of $N$ particles governed by
Hamilton's equation of classical mechanics

$$
\dot{\mbox{\boldmath $q$}}\ =\ {{\partial H}\over{\partial \mbox{\boldmath
$p$}}}\ ,  \qquad
\dot{\mbox{\boldmath $p$}}\ =\ -\ {{\partial H}\over{\partial \mbox{\boldmath
$q$}}}\ , \eqno(2.1)$$
\noindent
where ({\boldmath $q, p$}) are the positions and momenta of the particles. We
assume that these particles are contained in a rectangular domain of volume
$V$. At the borders, we may consider either hard walls of infinite mass or
periodic boundary conditions. In the latter case, the total momentum is
conserved in addition to the total energy. Moreover, to simplify the relation
with ergodic theory, we use the microcanonical ensemble and work on energy
shells
$H = E$.

In the large system limit $(N,V \rightarrow \infty$ with $N/V = n$),
irreversible processes in such a classical many-body
system may be described by hydrodynamic equations such as the Navier-Stokes
equations, or the diffusion equation, the chemical kinetic equations ruling
the time evolution of chemical concentrations, or the equations of
electrical conductivity which incorporate Ohm's law. These phenomenological
equations contain dissipative terms which are dependent on transport and
rate coefficients. The aim of nonequilibrium statistical mechanics is to
obtain these coefficients in terms of the microscopic Hamiltonian
equations. Since the work of Maxwell and Boltzmann, several methods have
been developed to calculate these coefficients. The most general method,
developed by Green and Kubo \cite{Gre50,Kub57}, is the time correlation
function
approach whereby the transport and rate coefficients, are given as
time-integrals
of autocorrelations of the fluxes,

$$\alpha \ = \ \int_0^{\infty} \ \lim_{V\to\infty} \ \langle J_0^{(\alpha)} \
J_t^{(\alpha)} \rangle \ dt \ , \eqno(2.2)$$

\noindent
where $J^{(\alpha )}_{t}$ is the flux at time $t$ corresponding to the
coefficient $\alpha$. It is a function of the canonical variables ({\boldmath
$q,p$}) and is obtained by solving the equations of motion for a time $t$ after
an
initial time so that

$$J^{(\alpha )}_{t} (\hbox{\boldmath $q,p$}) = J^{(\alpha )}_{0}
\lbrack\Phi^t (\hbox{\boldmath ${q},p$)}\rbrack\ , \eqno(2.3)$$

\noindent
where $\Phi^t$ denotes the flow in phase space induced by Hamiltonian
equations (2.1). We can express this in terms of an $N$-particle streaming
operator

$$J_t^{(\alpha)} \ = \ \exp(- t \hat L) \ J_0^{(\alpha)} \ , \eqno(2.4)$$

\noindent
with the Liouvillian operator given in terms of a Poisson bracket
expression $\hat{L} = \lbrace H, \cdot \rbrace$. The average
$\langle \cdot \rangle$ in equation (2.3) is taken over a microcanonical
ensemble in our case.

It will be useful for us to apply the formulation of the Green-Kubo
expressions, as obtained by moments $G^{(\alpha )}_{t}$ such that
the fluxes $J^{(\alpha )}_{t}$ are derivatives, as \cite{Hel60}

$$J_t^{(\alpha)} \ = \ {{d}\over{dt}} \ G_t^{(\alpha)} \ , \eqno(2.5)$$

\noindent
An integration by parts shows that

$$\langle ( G_t \ - \ G_0)^2 \rangle \ = \ \int_0^t \ \int_0^t \ \langle J_{t'}
J_{t''} \rangle \ dt' \ dt'' \ = \ 2 \ t \ \int_0^t \ \biggl( 1 \ - \
{{\tau}\over t}\biggr) \ \langle J_0 J_{\tau} \rangle \ d\tau \ . \eqno(2.6)$$

\noindent
Accordingly, if the following condition is satisfied

$$\lim_{t\to \infty} \ {1 \over t} \int_0^t \ \tau \ \langle J_0 J_{\tau}
\rangle \ d\tau \ = \ 0 \ , \eqno(2.7)$$

\noindent
we obtain the equality

$$\lim_{t\to\infty} \ {{\langle ( G_t \ - \ G_0 )^2\rangle}\over{2t}} \ = \
\int_0^{\infty} \ \langle J_0 J_{\tau} \rangle \ d\tau \ . \eqno(2.8)$$

\noindent
As a consequence, the transport and rate coefficients are also given by
\cite{Hel60}

$$\alpha \ = \ \lim_{t\to\infty} \ {1 \over{2t}} \ \lim_{V\to\infty} \ \langle
\ \Bigl\lbrack G_t^{(\alpha)} \ - \ G_0^{(\alpha)} \Bigr\rbrack^2 \rangle \ .
\eqno(2.9)$$

\vskip 1 cm
\hrule \vskip1pt \hrule height1pt
\vskip 0.1 cm

\begin{center}
{\bf Table I. Helfand's moments.}
\end{center}

\vskip 0.1 cm

$$
\vcenter{\openup1\jot \halign{#\hfil&\qquad#\hfil\cr
  {\it process}  &  {\it moment}  \cr
 self-diffusion  & $G^{\rm (D)} \ = \ x_i$ \cr
 shear viscosity & $G^{(\eta)} \ = \ {1 \over{\sqrt{Vk_{\rm B}T}}}
\ \sum_{i=1}^{N} \ x_i \ p_{iy} $ \cr
 bulk viscosity ($\psi = \zeta + {4 \over 3} \eta$) & $G^{(\psi)} \ = \ {1
\over{\sqrt{Vk_{\rm B}T}}}\ \sum_{i=1}^{N} \ x_i \ p_{ix} $ \cr
 heat conductivity & $G^{(\kappa)} \ = \ {1 \over{\sqrt{Vk_{\rm B}T^2}}}
\ \sum_{i=1}^{N} \ x_i \ (E_i \ - \ \langle E_i \rangle)$ \cr
 charge conductivity & $G^{\rm (e)} \ = \ {1 \over{\sqrt{Vk_{\rm B}T}}}
\ \sum_{i=1}^{N} \ eZ_i \ x_i$ \cr
 chemical reaction rate & $G^{\rm (r)} \ = \ {1 \over{\sqrt{Vk_{\rm B}T}}}
\ ( N^{(\rm r)} \ - \ \langle N^{(\rm r)} \rangle )$\cr}}$$

\vskip 0.1 cm
\hrule height1pt \vskip1pt \hrule
\vskip 0.5 cm

In  Table I we list moments appropriate for each transport or rate
coefficient, where $E_i = {{p^{2}_{i}}\over{2m}} + {1 \over 2} \sum_{j(\neq i)}
V_{ij}$ is the energy of particle $i$ and $V_{ij}$ is the potential energy
of interaction between particles $i$ and $j$. Appendix A describes in more
detail the case of chemical reactions.

Eq. (2.9) shows that, in the case the transport or rate coefficients are
well defined (i.e. are positive and finite), Helfand's moments undergo a
diffusive type of motion along the axis of the moment $G^{(\alpha )}$.
Therefore, the moment may be considered as a random variable having a
probability density $p(g)$ obeying a diffusion-type equation, in an
equilibrium ensemble

$${{\partial p}\over{\partial t}} \ = \ \alpha \ {{\partial^2 p}\over{\partial
g^2}} \ , \eqno(2.10)$$

\noindent
with the transport or rate coefficient $\alpha$ a diffusion coefficient.
This equation is the Fokker-Planck equation governing the equilibrium
fluctuations of the Helfand moments. Equivalently, one can recover Eq. (2.9) by
supposing that the moments satisfy a Langevin stochastic differential equation

$$\dot G^{(\alpha)} \ = \ J_t^{(\alpha)} \ , \eqno(2.11)$$

\noindent
where the flux $J^{(\alpha )}_{t}$ is a white noise

$$\langle J_t^{(\alpha)} \rangle \ = \ 0 \ , \qquad \langle J_0^{(\alpha)}
J_t^{(\alpha)} \rangle \ = \ 2 \alpha \ \delta(t) \ . \eqno(2.12)$$

\noindent

It is important to note that equations (2.10) or (2.12) are to be considered
as a simple representation of the results of the time correlation function
method for time scales which are much longer than the time necessary for the
time correlation functions of the microscopic currents to decay to zero.
We suppose that this approach applies to situations where, for example, long
time tails in the correlation functions decay sufficiently rapidly for the
transport coefficients as defined by Eq. (2.2) to exist. In this way we can
still treat processes which are diffusive on long time scales, but have
correlations on shorter time scales. Accordingly, the existence of a flux
autocorrelation function which differs from a delta distribution on short
time scales is still perfectly compatible with the validity of the
diffusion-type equation, (2.10), on long time scales.

In this discussion we have tacitly assumed that we consider a physical
system in the proper thermodynamic limit. However, for systems of finite
volume, there is an upper limit on the times we can consider because the
ranges of variation of the Helfand moments are bounded. For example, if the
system consists of hard spheres placed in a cubical box of length $L$ on a
side, the positions of each particle vary only in the interval

$$- \ {L \over 2} \ \leq \ x_i , \ y_i , \ z_i \ \leq \ + \ {L \over 2} \ ,
\eqno(2.13)$$

\noindent
while the momenta can only take values in the interval

$$- \ \sqrt{2mE} \ \leq \ p_{ix} , \ p_{iy} , \ p_{iz} \ \leq \ + \
\sqrt{2mE} \ , \eqno(2.14)$$

\noindent
where $E = E_{\rm tot} = \frac{3}{2} N k_{\rm B} T$ is the total energy.
Similarly, the energy of a particle can only lie in the range

$$0 \ \leq \ E_i\ \leq \ {3 \over 2} \
k_{\rm B} T \ N \ . \eqno(2.15)$$

\noindent
Because of these bounds, the moments are always of bounded variation in the
interval

$$\vert G^{(\alpha)} \vert \ \leq \ C\ N^{\delta^{(\alpha)}} \ ,
\eqno(2.16)$$

\noindent
where $C$ is some constant, and $\delta^{(\alpha)}$ are positive exponents
which are respectively $\delta^{\rm (D)} = 1/3 ,\ \delta^{(\eta )} = 4/3 ,\
\delta^{(\psi )} = 4/3 ,\ \delta^{(\kappa )} = 11/6 ,\ \delta^{\rm (e)} =
5/6$, and $\delta^{\rm (r)} = 1/2$, for three-dimensional gases. Accordingly,
the range of variation of the moments grow with the size of the system,
when the density and other intensive variables are kept constant. Of course,
most variations of the Helfand moments will be due to microscopic motions of
the particles and thus will be much smaller than the bounds in Eq. (2.16).
At any rate, as the system gets larger, we expect that the diffusive-like
behavior of the moments, described by  Eq. (2.10), will be valid over
increasingly larger regions of variations of the $G^{(\alpha)}$.

After the discussion about the limited range over which the Helfand moments
obey the diffusive-type equation (2.10) for finite systems, we may proceed.

\section{Escape-Rate Formalism}

Within the range of validity of (2.10), we may set up a problem of first
passage for the moment $G^{(\alpha )}$ corresponding to the transport or
rate processes of coefficient $\alpha$. We consider a statistical ensemble
formed
by copies of the system which we assume to be at equilibrium and
microcaninical at total energy $E$ (and eventually at fixed total momentum in
the case of periodic boundary conditions: $\mbox{\boldmath $P$}_{\rm tot} =
\sum^{N}_{i=1} \mbox{\boldmath $p$}_i$). For each copy, the motion of the
Hamiltonian system (2) is integrated from the initial conditions and the
Helfand moment is calculated along the trajectory. At each time, we count the
number of copies, ${\cal N}^{(\alpha )}(t)$, for which the moment is still in
the
following interval

$$- \ {{\chi} \over 2} \ \leq \ G_t^{(\alpha)} \ \leq \ + \ {{\chi} \over 2} \
,
\eqno(3.1)$$

\noindent
where the size of the interval $\chi$ is sufficiently large to be in the
regime of diffusion of the moment but not too large with respect to the total
variation interval of the moment allowed by the finiteness of the system. We
are
here defining a problem of first passage which can be solved using the
eigenvalues and eigenfunctions of the Fokker-Planck equation (2.10) with the
boundary conditions

$$p(-{\chi}/2) \ = \ p(+{\chi}/2) \ = \ 0 \ . \eqno(3.2)$$

\noindent
The solution of this eigenvalue problem is well known to be

$$p(g,t) \ = \ \sum_{n=1}^{\infty} \ c_n \ \exp(-\gamma_n^{(\alpha)} t) \
\sin\biggl( {{\pi n}\over {\chi}} \ g \biggr) \ , \qquad \hbox{with} \qquad
\gamma_n^{(\alpha)} \ = \ \alpha \ \biggl({{\pi n}\over {\chi}} \biggr)^2 \ ,
\eqno(3.3)$$

\noindent
where the constants $c_n$ are fixed from the initial probability density
$p(g,0)$. The number of copies of the statistical ensemble which are still in
the interval (3.1) is then given by

$${\cal N}^{(\alpha)}(t) \ = \ \int_{-\chi/2}^{+\chi/2} \ p(g,t) \ dg \ .
\eqno(3.4)$$

\noindent
At long times, the decay is dominated by the slowest decay mode
corresponding to the smallest decay rate $\gamma^{(\alpha )}_{1}$ which
defines the escape rate of the moment out of the interval (3.1),

$$\gamma_1^{(\alpha)} \ = \ \alpha \ \biggl( {{\pi}\over {\chi}}\biggr)^2 \ .
\eqno(3.5)$$

We are now in position to establish a relationship with the deterministic
dynamics. The Hamiltonian classical motion of the many-body system is
chaotic in many cases. This property has been proved by Sinai and coworkers
for some simple hard sphere gas models \cite{Sin70,Sin79}. Also
strong numerical evidence exists which shows that half of the Lyapunov
exponents
are typically positive in systems of statistical mechanics like the
Lennard-Jones
gas at room temperatures \cite{PosHoo88}.	We suppose that the decomposition of
phase space into  ergodic components is understood and that, beside the
decomposition on the  known constants of motion, there is only a single main
ergodic component.

We consider  the set of all the trajectories for which the Helfand moment
remains forever within the interval (3.1). Because most of the
trajectories are expected to exit this interval, the trapped trajectories
must be exceptional and highly unstable forming a set of measure zero with
respect to the microcanonical probability measure. Based on earlier work on
diffusion in the Lorentz gas and in related models and on the basis that
the trajectories are typically of saddle type in systems of statistical
mechanics \cite{GasNic90,GasBar91,Gas92,GasRic89,Gas94}, we assume that the set
of
trajectories is a fractal repeller. Indeed, this set is of  vanishing measure
but
may still contain an uncountable infinity of periodic and nonperiodic
trajectories. A set satisfying these conditions is necessarily a fractal
\cite{EckRue85}. Moreover, it is composed of unstable trajectories of saddle
type
so that it forms a repeller (of saddle type) in phase space.  These properties
can
be proved for particular models like the multibaker area-preserving map
\cite{Gas92} as well as the array of disk scatterers composing the  periodic
Lorentz gas. Accordingly, it seems reasonable to assume that, for more general
systems, such as a gas of hard spheres the set of trajectories for which the
moments satisfy Eq. (3.1), form a fractal repeller, with properties to be
described in the next paragraph.

A fractal repeller is characterized by different quantities and, especially,
by an escape rate which is the deterministic analogue of the escape rate
obtained
in the preceding first-passage problem. Moreover, in chaotic systems, the
escape rate is related to the sum of positive Lyapunov exponents
minus the Kolmogorov-Sinai entropy per unit time if these quantities are
well defined and positive \cite{EckRue85,KanGra85,BohRan87,Tel,GasRic89}. These
quantities are evaluated for the natural invariant probability measure whose
support is the fractal repeller. For the natural invariant measure, each cell
of
phase space has a weight which is inversely proportional to the local Lyapunov
numbers (stretching factors).  Therefore we have
\cite{EckRue85,KanGra85,BohRan87,Tel,GreOttYor88,Gas92,GasRic89}

$$\gamma_1^{(\alpha)} \ = \ \sum_{\lambda_i > 0} \ \lambda_i({\cal
F}_{\chi}^{(\alpha)}) \ - \ h_{\rm KS}({\cal F}_{\chi}^{(\alpha)}) \ ,
\eqno(3.6)$$

\noindent
where we denote by ${\cal F}^{(\alpha )}_{\chi}$ the fractal repeller
formed by the trapped trajectories for which the Helfand moment
$G^{(\alpha)}_{t}$ remains forever in the interval (3.1).

Combining the deterministic result (3.6) with the statistical result (3.5),
we obtain the relationship

$$\alpha \ = \ \lim_{{\chi} \to \infty} \ \biggl( {{\chi} \over{\pi}} \biggr)^2
\ \lim_{V\to\infty} \ \Biggl\lbrack \sum_{\lambda_i > 0} \ \lambda_i({\cal
F}_{\chi}^{(\alpha)}) \ - \ h_{\rm KS}({\cal F}_{\chi}^{(\alpha)})\Biggr\rbrack
\ , \eqno(3.7)$$

\noindent
where the limit $V\rightarrow \infty$ denotes the thermodynamic limit to
be taken before the limit $\chi\rightarrow \infty$ which is internal to the
system.

With Table I, Eq. (3.7) shows how a general transport or rate coefficient
can in principle be related to the Lyapunov exponents and the Kolmogorov-Sinai
entropy of a fractal repeller. This fractal repeller is the phase space
object corresponding to the escape process of the Helfand moment associated to
the transport or rate coefficient. In this way, a connection is established
between statistical and mechanical considerations in phase space.

A remark is now in order about the magnitude of the quantities appearing in
(3.7).  The sum of positive Lyapunov exponents and the KS-entropy per
unit time are very large, of the order of the number of particles times the
inverse of a typical kinetic time scale \cite{GasWan93}.  On the other hand,
the
escape rate, which is the difference between two such large numbers, has a much
smaller magnitude given by the time scales characteristic of hydrodynamics.  In
this way, the kinetic and hydrodynamic levels are naturally connected with a
formula like (3.7).

\section{Conclusion}

We have shown that all of the transport coefficients for a simple fluid
and that chemical reaction rate coefficients can be expressed in terms of an
escape rate from an appropriate fractal repeller. This completes a line of
argument initiated by previous work on the coefficient of diffusion for a
particle moving in a periodic Lorentz system \cite{GasNic90}, and for a
multibaker
map \cite{Gas92}.  In a separate paper \cite{GasJRD94}, we will show how the
escape-rate formula can be applied to several classes of dynamical systems like
the hard sphere gas and the lattice gas automata.  We also describe there a
large
deviation formalism which allows us to extend the application of the escape
rate
formula (3.7) from hyperbolic to nonhyperbolic systems, in particular, using
the Ruelle pressure function.  It now remains to apply this formalism to a
number of examples in order to understand the  physical and mathematical
consequences of this approach to transport in fluid systems.  Applications of
this formalism to certain types of one-dimensional diffusion problems will be
presented in \cite{KlaJRD94}, and applications to Lorentz lattice gas cellular
automata will be presented in \cite{ErnJRDJac94}.  Many further applications
are possible.

Many interesting problems remain open in the present context, in particular:
\begin{itemize}
\item[a)] To provide experimental evidence for the microscopic chaos at the
basis of the present theory.  Suggestions along this line has been discussed
elsewhere \cite{GasCanWe}.
\item[b)] To provide a more rigorous mathematical derivation of the escape-rate
formalism used here.
\item[c)] To establish the connection between this formalism and those based
upon period-orbit theory \cite{CviEckGas94} and upon a study of
eigendistributions
of the Liouville operator, and Ruelle resonances
\cite{Gas92,GasHasDri92,Tas94}.
Already, such a connection has been discussed elsewhere \cite{Gas94} where it
was shown for the multibaker model that the diffusive-like eigendistributions
of
the Liouville operator shares self-similar properties with the fractal repeller
underlying diffusion.  Moreover, both approaches make use of the leading
eigenvalues (3.5) of the Liouville operator and of its dependence on the
wavenumber of the diffusive eigenmode ($k=\pi/L$).
\item[d)] To extend this
formalism to include transport in mixtures, and to a treatment of higher order,
and nonlinear transport processes.
\item[e)] To study the structure of the
fractal repeller for interesting cases where, as a consequence of long time
tail
effects, transport processes are anomalous, such as two dimensional fluid
systems \cite{Tail}.
\item[f)] To establish the connection between the
escape-rate formalism for transport and the Gaussian thermostat formalism of
Hoover and Posch \cite{HooPos87}, and of Evans, Morris, Cohen, and coworkers
\cite{EvaCohMor90,BarEvaCoh93}.  The Gaussian thermostatted systems belong to
a class of dynamical systems which are time reversible without being
volume preserving.  The characteristic quantities describing chaos in these
thermostatted systems have been described in \cite{CheEyiLebSin93,EvaCohMor93}.

\end{itemize}

\vfill\eject

\noindent
{\bf Acknowledgements.} J.R.D. would like to express his appreciation to
Celia Shapiro, and to his physicians Dr. Kenneth Goldstein and Dr. Melvin
Stern for helping him through a difficult illness. He would like to thank his
many
friends and colleagues, especially Profs. E.G.D. Cohen, J. Dufty,
M.H. Ernst, C. Grebogi, T.R. Kirkpatrick, E. Ott, J.V. Sengers, D. Thirumulai,
J. Yorke, his coauthor, C. Beck, and Mr. R. Klages for their considerable help,
advice, and support which assisted the resumption of his scientific career.
He would also like to thank the Institute for Theoretical Physics, University
of Utrecht, for its warm hospitality during the spring semester of 1994
when this work was completed.  P.G. is grateful to Prof. G. Nicolis for support
in
this research and as well to the National Fund for Scientific Research (F. N.
R.
S. Belgium) for financial support.

\vfill\eject

\centerline{\Large{\bf Appendix A: The Case of Chemical Reactions}}

\noindent{\bf A.1. Summary of thermodynamic results}

For a chemical reaction like

$$\sum_{\gamma=1}^c \ \nu_{\gamma}^+ \ X_{\gamma} \ \leftrightarrow \
\sum_{\gamma=1}^c \ \nu_{\gamma}^- \ X_{\gamma} \ , \eqno({\rm A}.1)$$

\noindent
the numbers of particles of the reactants and products are changing at each
step
of the reaction according to

$${{\Delta N_1}\over{\nu_1}} \ = \ {{\Delta N_2}\over{\nu_2}} \ = \ \dots \ = \
{{\Delta N_c}\over{\nu_c}} \ = \ \Delta N^{(\rm r)} \ , \eqno({\rm A}.2)$$

\noindent
where $\nu_{\gamma} = \nu_{\gamma}^- - \nu_{\gamma}^+$ are the stoichiometric
coefficients \cite{GlaPri71}.  The degree of progress of the reaction can be
measured in terms of the variation of the number $N^{(\rm r)}$ characterizing
the
reaction.  We also introduce the chemical concentrations,
$C_{\gamma}=N_{\gamma}/V$.  At the phenomenological level of thermodynamics,
the velocity of the reaction is defined by

$$\dot{\bar C}^{(\rm r)} \ = \ {{\dot{\bar C}_{\gamma}}\over{\nu_{\gamma}}} \ =
\ w \ , \eqno({\rm A}.3)$$

\noindent
where $\bar C_{\gamma}$ are the average chemical concentrations
\cite{GlaPri71}.  The dependence of the reaction velocity on the concentrations
themselves is given by the mass action law \cite{GlaPri71,NicPri77}

$$w \ =\ k_+ \ \prod_{\gamma=1}^c \ C_{\gamma}^{\nu_{\gamma}^+} \ - \ k_- \
\prod_{\gamma=1}^c \ C_{\gamma}^{\nu_{\gamma}^-} \ . \eqno({\rm A}.4)$$

\noindent
The affinity of the reaction is defined by

$$A \ = \ - \ \sum_{\gamma=1}^c \ \nu_{\gamma} \ \mu_{\gamma} \ = \ - \ k_{\rm
B} T \ \ln \ \prod_{\gamma=1}^c \ \biggl( {{C_{\gamma}}\over{C_{\gamma}^{\rm
eq}}} \biggr)^{\nu_{\gamma}} \ , \eqno({\rm A}.5)$$

\noindent
where $\mu_{\gamma}=\mu_{\gamma}^0 + k_{\rm B} T \ln C_{\gamma}$ is the
chemical potential of the species $\gamma$ and it is known that the
affinity vanishes at thermodynamic equilibrium: $A^{\rm eq} = 0$
\cite{GlaPri71,NicPri77}.  Near the thermodynamic equilibrium, both the
reaction
velocity $w$ and the affinity $A$ can be expanded in terms of the variations of
the chemical concentrations around their equilibrium values, $C_{\gamma} =
C_{\gamma}^{\rm eq} + \Delta C_{\gamma}$.  In the linear approximation, we
obtain
the equality

$$w \ \simeq \ w_+^{\rm eq} \ {A \over{k_{\rm B} T}} \ , \qquad \hbox{with}
\qquad w_+^{\rm eq} \ = \ k_+ \ \prod_{\gamma=1}^c \ C_{\gamma}^{\rm eq} \ ,
\eqno({\rm A}.6)$$

\noindent
which allows us to obtain the Onsager coefficient of this chemical process

$$w \ \simeq  \ {{LA} \over{T}} \ , \qquad \hbox{with} \qquad
L \ = \ {{w_+^{\rm eq}}\over{k_{\rm B}}}\ . \eqno({\rm A}.7)$$

\noindent
Yamamoto \cite{Yam60}and, later, Zwanzig \cite{Zwa65} have shown that the
Onsager
coefficient of the chemical reaction is given by the following integral of the
autocorrelation function in a classical system

$$L \ = \ {1 \over{Vk_{\rm B}}} \ \int_0^{\infty} \ \langle \dot N_0^{(\rm r)}
\dot N_t^{(\rm r)} \rangle \ dt \ , \eqno({\rm A}.8)$$

\noindent
where the corresponding flux is here proportional to the time derivative of the
number $N^{(\rm r)}$ of particles which is characteristic of the reaction.
Accordingly, we obtain the result given in Table I for the reaction coefficient

$$\alpha \ = \ {w \over A} \ = \ {L \over T} \ , \eqno({\rm A}.9)$$

\noindent
with $\langle N^{(\rm r)}\rangle =N^{(\rm r), eq}$.

\vfill\eject

\noindent{\bf A.2. Master equation approach}

The escape-rate formula can be obtained for chemical reactions by using the
master
equation approach \cite{NicPri77}.  As an example, we consider the
isomerization

$$A \ \leftrightarrow \ B \ . \eqno({\rm A}.10)$$

\noindent
The numbers of particles $A$ and $B$ are the random variables of this
process.  The total number of particles is conserved, $N_A + N_B = N = N_{\rm
tot}$, so that the process is completely determined by the knowledge of the
number
of particles $A$.  The evolution equation of the probability $P(N_A)$ that the
system contains $N_A$ particles is \cite{NicPri77}

$${d \over{dt}} P(N_A) \ = \ k_+ (N_A+1) \ P(N_A+1) \ + \ k_- (N -
N_A+1) \ P(N_A-1) $$
$$- \ k_+ N_A \ P(N_A) \ - \ k_- (N - N_A) \ P(N_A)\ . \eqno({\rm A}.11)$$

\noindent
Introducing the fraction $0 \leq f=N_A/N \leq 1 $ of particles $A$
and defining a probability density which is function of this continuous
concentration according to $p(f)=p(N_A/N)=P(N_A)$, we use Taylor
expansions in the master equation (A.11) to obtain the Fokker-Planck
equation

$${{\partial p}\over{\partial t}} \ + \ {{\partial}\over{\partial f}} \Bigl (
\dot f \ p \Bigr) \ = \ {{\partial}\over{\partial f}}\Biggl\lbrack \ D(f) \
{{\partial p}\over{\partial f}}\Biggr\rbrack \ , \eqno({\rm A}.12)$$

\noindent
where

$$\dot f \ = \ -\ k_+ \ f \ + \ k_- (1-f) \ + \ {\cal O}(1/N) \ , \eqno({\rm
A}.13)$$

\noindent
is the macroscopic rate equation while

$$D(f) \ = \ {1 \over {2N}} \ \Bigl\lbrack k_+ f \ + \ k_-(1-f) \Bigr\rbrack \
,
\eqno({\rm A}.14)$$

\noindent
is a concentration-dependent diffusion coefficient.  This Fokker-Planck
equation
determines the time evolution of the probability density of a
stochastic process in which the amplitude of the noise is proportional to
$1/\sqrt{N}$.  Let us consider these thermodynamic fluctuations around the
equilibrium concentration

$$f^{\rm eq} \ = \ {{k_-}\over{k_+ + k_-}} \ , \eqno({\rm A}.15)$$

\noindent
in the vicinity of which the Fokker-Planck equation becomes a diffusion
equation

$${{\partial p}\over{\partial t}} \ = \ D(f^{\rm eq}) \
{{\partial^2 p}\over{\partial f^2}} \ , \eqno({\rm A}.16)$$

\noindent
with

$$D(f^{\rm eq}) \ = \ {{k_+ k_-}\over{N(k_+ + k_-)}} \ . \eqno({\rm A}.17)$$

\noindent
On this ground, we can apply the argument of first passage for the Helfand
moment
associated with the chemical reaction, which is given in Table I

$$G^{(\rm r)} \ = \ {1 \over{\sqrt{Vk_{\rm B} T}}} \ ( N_A \ - \ N_A^{\rm eq})
\
. \eqno({\rm A}.18)$$

\noindent
The diffusion coefficient (A.17) shows that the moment $G^{\rm (r)}$ is a
random variable of Brownian type in a small neighborhood of the
thermodynamic equilibrium.  Indeed, the moment is related to the fraction
$f=N_A/N$ of particles $A$ according to

$$G^{(\rm r)} \ = \ {N \over{\sqrt{Vk_{\rm B} T}}} \ ( f \ - \ f^{\rm eq}) \
. \eqno({\rm A}.19)$$

\noindent
We are looking for the first time at which the moment reaches the boundaries of
the interval

$$- \ {{\chi} \over 2} \ < \ G^{\rm (r)} \ < \ + \ {{\chi} \over 2} \ .
\eqno({\rm A}.20)$$

\noindent
This first-passage problem corresponds to the first escape of the fraction $f$
out of the interval

$$f^{\rm eq} \ - \ {{\varepsilon} \over 2} \ < \ f \ < \ f^{\rm eq} \ + \
{{\varepsilon} \over 2} \ , \eqno({\rm A}.21)$$

\noindent
with

$$\varepsilon \ = \ {{\sqrt{Vk_{\rm B} T}}\over N} \ {\chi} \ . \eqno({\rm
A}.22)$$

\noindent

We consider a statistical ensemble formed by copies of the system and we count
the number of copies which remain within the interval (A.20) at time $t$.
Since the moment $G^{\rm (r)}$ or the fraction $f$ follows a diffusion process,
the number of copies is decaying exponentially

$${\cal N}^{\rm (r)} (t) \ \sim \ \exp\Bigl\lbrack-\gamma_1^{\rm (r)}
t\Bigr\rbrack \ , \eqno({\rm A}.23)$$

\noindent
at a rate $\gamma_1^{\rm (r)}$ given by

$$\gamma_1^{\rm (r)} \ \simeq \ D(f^{\rm eq}) \ \biggl(
{{\pi}\over{\varepsilon}} \biggr)^2 \ , \eqno({\rm A}.24)$$

\noindent
in terms of the diffusion coefficient (A.17) of the
thermodynamic fluctuations around equilibrium.  We mention that this
exponential decay is the slowest decay dominating the time evolution at long
times after the faster decay modes have died out.  We can replace the size
$\varepsilon$ (A.22) of the escape interval of the variable $f$ by the
size ${\chi}$ of the escape interval of the corresponding Helfand moment
$G^{\rm (r)}$ and we get

$$\gamma_1^{\rm (r)} \ \simeq \ \alpha \ \biggl( {{\pi}\over{\chi}}
\biggr)^2 \ , \eqno({\rm A}.25)$$

\noindent
with the rate coefficient

$$\alpha \ = \ {{N k_+ k_-}\over{V k_{\rm B} T (k_+ + k_-)}} \ = \ {w \over A}
\ , \eqno({\rm A}.26)$$

\noindent
so that we recover the rate coefficient (A.7) and (A.9) of the macroscopic
theory since

$$w \ = \ {{w_+^{\rm eq}}\over {k_{\rm B} T}} \ , \qquad \hbox{with} \qquad
w_+^{\rm eq} \ = \ k_+ \ C_A^{\rm eq} \ = \ {{N k_+ k_-}\over{V(k_+ + k_-)}}\
. \eqno({\rm A}.27)$$

\noindent
In this way, we see the consistency of the first-passage problem applied to
the thermodynamic fluctuations described by the master equation with the
macroscopic theory as well as with the Green-Kubo formula.  The relation to
the fractal repeller of the deterministic dynamics is discussed in full
generality in the main part.

\end{document}